\documentclass[acmtog]{acmart}
\acmSubmissionID{1397s3}

\usepackage{booktabs} 

\usepackage{amsmath}		
\usepackage{float}
\usepackage{color}
\usepackage{enumitem}
\usepackage{tabularray}
\usepackage{colortbl}
\usepackage{xcolor}

\usepackage{algorithm}
\usepackage{algpseudocode}
\newcommand{\pass}[1]{#1}
\newcommand{\revise}[1]{#1}

\acmJournal{TOG}

\copyrightyear{2025}
\acmYear{2025}
\setcopyright{none}
\setcctype{by}
\acmConference[SIGGRAPH Conference Papers '25]{Special Interest Group on Computer Graphics and Interactive Techniques Conference Conference Papers }{August 10--14, 2025}{Vancouver, BC, Canada}
\acmBooktitle{Special Interest Group on Computer Graphics and Interactive Techniques Conference Conference Papers (SIGGRAPH Conference Papers '25), August 10--14, 2025, Vancouver, BC, Canada}
\acmDOI{10.1145/3721238.3730757}
\acmISBN{979-8-4007-1540-2/2025/08}

\newcommand{\timebracket}[1]{\{#1\}_{k=1}^T}
\newcommand{\SImage}{I_{k}^s}
\newcommand{\SImagecus}[1]{I_{#1}^s}
\newcommand{\SImagepredcus}[1]{\hat{I}_{#1}^s}

\newcommand{\TImage}{I_{k}^t}

\newcommand{\Scampose}{P_{k}^s}
\newcommand{\Scamposecus}[1]{P_{#1}^s}
\newcommand{\Tcampose}{P_{k}^t}

\newcommand{\SGaussian}{\cG_{k}^s}
\newcommand{\SGaussiancus}[1]{\cG_{#1}^s}
\newcommand{\SDepth}{\cD_{k}^s}
\newcommand{\framescale}{\alpha}
\newcommand{\camintrinsic}{K}
\newcommand{\stability}{\sigma_s}
\newcommand{\reconhead}{\bR}
\newcommand{\flow}[2]{F_{#1 \rightarrow #2}}
\newcommand{\camflow}[2]{F^{\text{cam}}_{#1 \rightarrow #2}}
\newcommand{\objectflow}[2]{F^{\text{obj}}_{#1 \rightarrow #2}}

\newcommand{\neighborimgs}{\cN_{k}}

\newcommand{\renderer}{\text{GSplat}}
\newcommand{\flowmodel}{\text{RAFT}}

\newcommand{\scaleLarge}{\tau_{\text{large}}}
\newcommand{\scaleDepth}{\tau_{\text{depth}}}

\newcommand{\firstcolor}[1]{\textbf{#1}}
\newcommand{\secondcolor}[1]{#1}

\newcommand{\winsize}{W}
\newcommand{\framesetsize}{S}

\newcommand{\bR}{\mathbf{R}}

\newcommand{\cD}{\mathcal{D}}

\newcommand{\cG}{\mathcal{G}}

\newcommand{\cL}{\mathcal{L}}
\newcommand{\cM}{\mathcal{M}}
\newcommand{\cN}{\mathcal{N}}

\newcommand{\figref}[1]{Fig.~\ref{#1}}
\newcommand{\secref}[1]{Section~\ref{#1}}

\newcommand{\tabref}[1]{Table~\ref{#1}}

\makeatletter
\DeclareRobustCommand\onedot{\futurelet\@let@token\@onedot}
\def\@onedot{\ifx\@let@token.\else.\null\fi\xspace}

\makeatother

\definecolor{darkgreen}{rgb}{0,0.7,0}
\definecolor{darkblue}{RGB}{31,119,180}
\definecolor{darkred}{RGB}{214,39,40}

\definecolor{bronze}{rgb}{1,1,0.6}
\definecolor{silve}{rgb}{0.969,0.796,0.600}
\definecolor{gold}{rgb}{0.941,0.592,0.600}

\definecolor{scene}{rgb}{0.745,0.486,0.451}
\definecolor{probe}{rgb}{0.767,0.767,0.767}
\definecolor{core_v}{rgb}{0.791,0.876,0.723}
\definecolor{core_m}{rgb}{0.918,0.705,0.541}
\definecolor{basis_m}{rgb}{0.647,0.760,0.889}

\begin{document}
\title[GaVS: 3D-Grounded Video Stabilization]{GaVS: 3D-Grounded Video Stabilization via Temporally-Consistent Local Reconstruction and Rendering}

\begin{teaserfigure}
    \centering
    \includegraphics[width=0.95\textwidth]{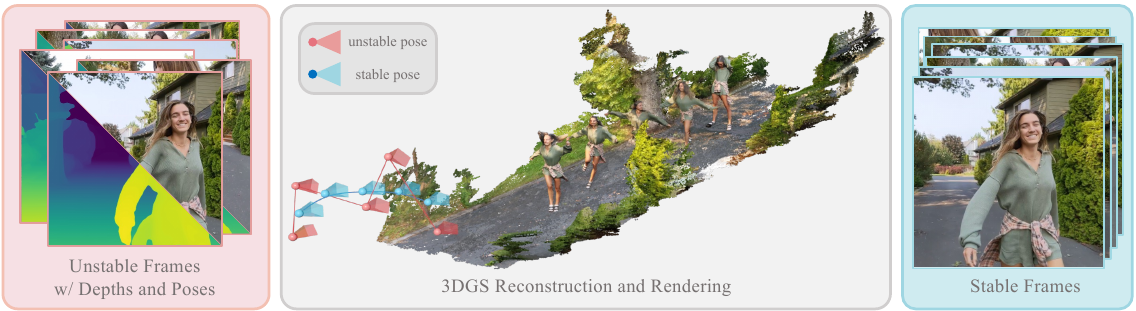}
    \caption{
    \pass{We study the task of video stabilization from a novel 3D perspective. Given an unstable video and 3D camera poses, we reconstruct in a feed-forward manner localized 3D scenes from the unstable poses in the form of Gaussian Splatting primitives, and render stabilized frames at smoothed poses. Our method is robust to diverse camera motions and scene dynamics, while ensuring full-frame stabilization and temporal consistency across local 3D reconstructions.}
    }
    \label{fig:teaser}
\end{teaserfigure}

\author{Zinuo You}
\authornote{\noindent Work partly done during Zinuo You's internship at Huawei Research Z\"urich. }
\affiliation{%
 \institution{ETH Z\"urich}
 \city{Z\"urich}
 \country{Switzerland}}
\email{zinyou@ethz.ch}
\author{Stamatios Georgoulis}
\affiliation{%
 \institution{Huawei Research Z\"urich}
 \city{Z\"urich}
 \country{Switzerland}
}
\email{stamatios.georgoulis@huawei.com}
\author{Anpei Chen}
\affiliation{%
\institution{ETH Z\"urich}
\city{Z\"urich}
\country{Switzerland}}
\email{chenanpei@westlake.edu.cn}

\author{Siyu Tang}
\affiliation{%
 \institution{ETH Z\"urich}
 \city{Z\"urich}
 \country{Switzerland}}
\email{siyu.tang@inf.ethz.ch}

\author{Dengxin Dai}
\affiliation{%
 \institution{Huawei Research Z\"urich}
 \city{Z\"urich}
 \country{Switzerland}
}
\email{dengxin.dai@huawei.com}

\begin{abstract}
Video stabilization is pivotal for video processing, as it removes unwanted shakiness while preserving the original user motion intent. Existing approaches, depending on the domain they operate, suffer from several issues (e.g. geometric distortions, excessive cropping, poor generalization) that degrade the user experience. 
To address these issues, we introduce \textbf{GaVS}, a novel 3D-grounded approach that reformulates video stabilization as a temporally-consistent `local reconstruction and rendering' paradigm. Given 3D camera poses, we augment a reconstruction model to predict Gaussian Splatting primitives, and finetune it at test-time, with multi-view dynamics-aware photometric supervision and cross-frame regularization, to produce temporally-consistent local reconstructions. The model are then used to render each stabilized frame. We utilize a scene extrapolation module to avoid frame cropping. 
Our method is evaluated on a repurposed dataset, instilled with 3D-grounded information, covering samples with diverse camera motions and scene dynamics. Quantitatively, our method is competitive with or superior to state-of-the-art 2D and 2.5D approaches in terms of conventional task metrics and new geometry consistency. Qualitatively, our method produces noticeably better results compared to alternatives, validated by the user study. Project Page: \href{https://sinoyou.github.io/gavs}{\textcolor{blue}{sinoyou.github.io/gavs}}.

\end{abstract}

%
%

%
%



\maketitle


\section{Introduction}

Nowadays, the task of video stabilization is the cornerstone of every video processing pipeline, as it brings many benefits. It has high effect in the perceived visual quality of a video. It allows the users to focus on the content. It reduces motion sickness. It enhances the professionalism of a video. It fuels applications ranging from cinema production, to drone footage, to action cameras, and many more. Conceptually, a video stabilization pipeline typically consists of three steps: (1) tracking the camera motion from the unstable video, (2) smoothing the camera motion to filter out unwanted shakiness while retaining the original motion intent of the user, and (3) synthesizing the stabilized video. In the literature the camera motion, and hence the 3D scene itself, is mainly modeled in the 2D domain, represented by frame transforms (e.g. affinities~\cite{grundmann2011auto}, homographies~\cite{bradley2021cinematic}) or optical flow (at the pixel~\cite{liu2014steadyflow} or grid~\cite{liu2013bundled,liu2016meshflow} level). However, the lack of explicit 3D modeling can easily result to unpleasant distortions in the stabilized video. While follow-up works introduced learning-based representations~\cite{wang2018deep,yu2018selfie,yu2020learning,zhang2023minimum}, depth cues~\cite{lee20213d}, or even input from additional sensors~\cite{liu2012video,shi2022deep,li2022deep} to alleviate distortions, their pretrained models may not generalize well to open domain videos with diverse camera motions and scene dynamics. Notably, all aforementioned works operate on a per-frame basis, i.e. they synthesize each stabilized frame individually by transforming its corresponding unstable frame, which inevitably leads to frame cropping, and hence loss of content for the users. While 2D or 2.5D multi-frame fusion~\cite{choi2020deep,liu2021hybrid,zhao2023fast,peng20243d} has been recently explored to achieve \textit{full-frame} (i.e. no cropping) video stabilization, these works are also prone to inconsistency and distortion artifacts, as they still lack explicit 3D scene modeling. 

In this paper, we go beyond existing works and introduce a novel 3D-grounded video stabilization pipeline, named \textbf{GaVS}, that minimizes inconsistency and distortion artifacts while achieving full-frame video stabilization without using additional sensors. It is built upon the Gaussian Splatting (GS)~\cite{kerbl20233d} scene representation, known for high-fidelity, yet efficient, novel view synthesis. In our pipeline, we reformulate video stabilization as a temporally-consistent `local reconstruction and rendering' problem, where we reconstruct the scene from the unstable video and render stabilized frames using smoothed camera trajectories. 

Our first key observation is that, in video stabilization the corresponding unstable and stabilized frames only differ slightly, making it unnecessary to globally reconstruct the scene for rendering the stabilized frames. Note that, global reconstruction is particularly challenging in our dynamic scenes. Therefore, we omit conventional global optimization, and propose to model the scene with a neural network that predicts local reconstructions in a feed-forward manner. In practice, this is achieved by enhancing a depth estimator with additional heads that predict per-pixel GS primitives. Compared to expensive global 4D optimization methods~\cite{wang2024shape,cho20244d}, this efficient design not only naturally unifies the reconstruction of static and dynamic scene parts, but also enables our pipeline to robustly model diverse and complex motions (e.g. human, vehicles and fluids motions) via per-frame dynamics in an implicit way. 

Our second key observation is that, in 3D-grounded video stabilization the minimization of artifacts inherently dictates consistency in neighboring local reconstructions. To achieve this, we propose a test-time optimization scheme that adapts the pretrained local reconstruction model for each unstable video. Specifically, each local reconstruction, originating from a single unstable frame, is rendered to neighboring unstable frames and supervised with a multi-view photometric loss\footnote{Note that, as our local reconstructions are parameterized by neural networks, this ensures that similar image patches from neighboring frames would have similar reconstructions. The neural network weights updated by a single gradient descent will simultaneously update the reconstructions from all the similar patches. Therefore, with this compact parametrization, our adaptation framework is very light and efficient, requiring only few minutes for optimizing a pretrained local reconstruction model.}. To account for dynamic objects that violate the local scene's epipolar geometry, we propose a flow-based compensation mechanism that generates pseudo ground truth for the multi-view supervision. Furthermore, we introduce regularization in the GS primitives to encourage consistency between neighboring local reconstructions. Finally, to ensure full-frame video stabilization, we extrapolate the input unstable frames with a state-of-the-art video completion model, before generating the local reconstructions.  

To evaluate the proposed pipeline, we repurpose the video stabilization dataset in~\cite{shi2022deep} for our 3D-grounded task, focusing on samples that cover a wide range of camera motions and scene content. We quantitatively and qualitatively compare against state-of-the-art video stabilization works, also introducing an evaluation based on the sparse and dense 3D reconstruction to verify the consistency of stabilized videos. GaVS demonstrates superior performance versus image-based full-frame 2D and 2.5D video stabilization works.
In summary, 

\begin{itemize}[left=2pt]
    \item We reformulate video stabilization as a novel 3D grounded scheme of local reconstruction and rendering, which is naturally robust to diverse camera motions and scene dynamics, is temporally-consistent, and is capable of full frame stabilization. 
    \item We propose a novel test-time optimization for each unstable video, leveraging multi-view dynamics-aware photometric supervision and cross-frame regularization, to achieve temporally-consistent reconstructions. To avoid frame cropping, we propose a scene extrapolation module, based on video completion.
    \item We provide a 3D-grounded dataset for our task by re-purposing an existing one, and introduce new metrics on the sparse and dense reconstruction to evaluate 3D scene consistency. Extensive experiments (quantitative, qualitative, user study) vs image-based and gyro-based methods demonstrate the merits of our method. 
\end{itemize}
\begin{figure*}[ht]
    \centering
    \includegraphics[width=0.92\textwidth]{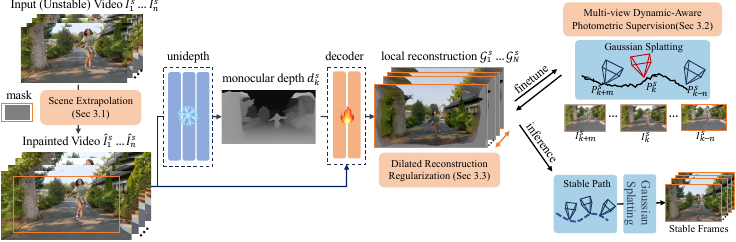}
    \caption{\textbf{\pass{Method Pipeline.}} We stabilize a video in two phases. 
    \textbf{1. Test-Time Finetuning.} Consider a pretrained reconstruction model that, first predicts depth from the input frame, then passes depth and the input frame to a decoder to obtain a local scene reconstruction in the form of Gaussian Splatting primitives. During the proposed test-time finetuning, we only update the decoder, which predicts position offsets and extra Gaussian Splatting attributes of point clouds. Local reconstructions are firstly extrapolated by video completion on the image domain, providing initial reconstruction (see \secref{sec:method-scene-extrapolation}). Then each local reconstruction is supervised on original images with a multi-view photometric loss to improve quality and reduce distortions. The loss contains compensation for dynamic objects (see \secref{sec:method-multiview-loss}). Furthermore, local reconstructions within a dilated temporal window are regularized by encouraging similarity in their primitives matched through dense correspondences from optical flow (see \secref{sec:method-window-regularization}).
    \textbf{2. Inference.} Each extrapolated reconstruction is rendered with its corresponding stabilized pose to get the stabilized full-frame video.}
    \label{fig:pipeline}
\end{figure*}

\section{Related Work}

As our approach is closely or loosely related to many diverse topics, like video stabilization, scene reconstruction, depth estimation, etc., in what follows we touch upon each topic. 
\\

\noindent \textbf{Video stabilization} 
In the literature, this task is mostly tackled in the 2D domain, by estimating the unstable camera trajectory, producing a stabilized camera trajectory, and using the latter to render stabilized frames. When it comes to modeling the unstable camera trajectory, a variety of representations have been employed, like affine transforms~\cite{grundmann2011auto}, homography transforms~\cite{bradley2021cinematic}, dense flow~\cite{liu2014steadyflow,yu2020learning,zhao2020pwstablenet}, mesh flow~\cite{liu2013bundled,liu2016meshflow,zhang2023minimum}, coded flow~\cite{liu2017codingflow}, feature tracks~\cite{lee2009video,goldstein2012video,koh2015video,yu2018selfie}, and even 6DoF poses~\cite{liu2023content}. To produce the stabilized camera trajectory, filter-based smoothing~\cite{goldstein2012video,liu2016meshflow}, regular optimization~\cite{liu2013bundled}, constrained optimization~\cite{grundmann2011auto,bradley2021cinematic}, and, as of recently, neural network~\cite{wang2018deep,yu2018selfie,yu2020learning,chen2021pixstabnet,xu2022dut,zhang2023minimum} approaches have been used. Finally, the rendering of stabilized frames follows a more standardized approach, where the employed representation (e.g. homography, mesh flow, dense flow, etc.) is first converted to a warp field which is then used to warp the unstable frame\footnote{Wherever applicable, effects like rolling shutter correction and Optical Image Stabilization offsets are taken into account during the rendering of stabilized frames.}. Unlike our 3D-grounded approach, these 2D stabilization works can easily result to unpleasant geometric distortions in the stabilized frames due to the lack of explicit 3D scene modeling.

To alleviate this issue, some works leveraged additional hardware, like depth cameras~\cite{liu2012video} and IMUs~\cite{shi2022deep,li2022deep}, or depth cues~\cite{lee20213d} to tackle video stabilization in the 2.5D or 3D domain. However, these works follow a static scene assumption, without explicit handling for dynamic objects as we do, which can still lead to geometric distortions on the surface of dynamic objects and along the boundaries between static and dynamic regions. In general, all aforementioned works suffer from two more issues. First, their pretrained models are prone to generalization issues when tested on open domain videos with diverse camera motions due the inherent domain gap. In contrast, our proposed test-time optimization scheme is designed to minimize this domain gap. Second, these works operate on a per-frame basis, i.e. they synthesize each stabilized frame individually by transforming its corresponding unstable frame. This inevitably leads to frame cropping, and hence loss of content for the users. Instead, our multi-view photometric supervision bundled with our extrapolation module achieve full-frame stabilization.

Similarly-minded full-frame stabilization works have been proposed in the literature. Here, the overarching scheme resolves around optical flow estimation, followed by multi-frame feature warping and fusion to produce stabilized frames with no cropping. Each approach adds its own touch, i.e. adopting a frame interpolation paradigm~\cite{choi2020deep}, employing hybrid fusion~\cite{liu2021hybrid}, using flow out-painting~\cite{zhao2023fast}, harnessing meta learning~\cite{ali2024harnessing}, or performing fusion in 3D~\cite{peng20243d}. However, due to the use of 2D cues (i.e. depth, optical flow) as a 3D proxy, these works do not model true 3D and inherit the limitations of 2D or 2.5D works: prone to geometric distortions, poor generalization to open domain videos. In contract, our `local reconstruction and rendering' paradigm with test-time adaptation introduces a novel 3D perspective for the task of video stabilization.
\\

\noindent \textbf{Scene reconstruction}
Recent advances in volumetric scene representations~\cite{mildenhall2021nerf,kerbl20233d} have significantly improved dense scene reconstruction for novel view synthesis. While several methods have been proposed to reconstruct both static~\cite{You2024CVPR, meuleman2023localrf} and dynamic~\cite{wang2024shape, Lei2024MoSca:Scaffolds, zhao2024pseudo, stearns2024dynamic} scenes from unconstrained camera trajectories, they often lack dynamic modeling or require lengthy optimization.
Our pipeline builds upon feed-forward variants that predict geometric and photometric information from a single or few views, typically modeled as multi-plane images~\cite{tulsiani2018layer,tucker2020single}, neural radiance fields~\cite{wang2021ibrnet,chibane2021stereo,chen2021mvsnerf,du2023learning,xu2024murf}, or, most relevant to ours, GS primitives~\cite{wewer24latentsplat,chen2024mvsplat}. A major drawback of few-view feed-forward reconstruction works is their static scene assumption. To remedy this situation, variants like Flash3D~\cite{szymanowicz2024flash3d} leverage monocular depth estimators to predict the GS primitives per view, implicitly handling dynamics along the way. However, such works still struggle to predict consistent outcomes in consecutive frames, due to the lack of explicit temporal (i.e. cross-view) modeling and inherent scale ambiguity. Our test-time optimization scheme with cross-frame regularization enhances consistency in consecutive frames reconstructions.
\\

\noindent \textbf{Video depth estimation}
Our work is loosely related to monocular depth estimation, which in a video setting, like ours, typically exhibits flickering and scale variance issues. The latter are detrimental for stabilization, which requires temporally smooth and consistent depth. To address this, test-time optimization approaches with flow-based losses~\cite{luo2020consistent,kopf2021robust} or learning-based neural modules~\cite{zhang2019exploiting,wang2023neural,shao2024learning} have been proposed. Recently, works like DepthCrafter~\cite{hu2024depthcrafter} and DepthAnyVideo~\cite{yang2024depth} investigated foundational models with temporal layers for consistent depth in open domain videos. However, these works only support a limited window of frames, still suffer from generalization issues, and most importantly only predict relative depth. In contrast, we employ a metric depth estimator~\cite{piccinelli2024unidepth} and leverage our test-time optimization scheme, with multi-view dynamics-aware photometric supervision and cross-frame regularization, to enforce temporally and geometrically consistent reconstructions.
\section{Method}

\pass{
Let us first introduce the notation. We use $\SImage$ to denote the \textit{source} frame from the input unstable video, and $\TImage$ for the \textit{target} stabilized frame. The goal of our stabilization pipeline is to generate a stabilized video $\timebracket{\TImage}$ with the same number of frames $T$ as the source video $\timebracket{\SImage}$, and without any cropping (i.e. no loss of content). For the camera we assume a pinhole model, parameterized by $\camintrinsic$, and use $\timebracket{\Scampose}$ and $\timebracket{\Tcampose}$ to denote the 3D poses of unstable frames and stabilized frames respectively. The stable poses are obtained via Gaussian filtering \revise{(see \secref{sec:optimization} for details).} 
}

Let us now motivate our approach. 
\pass{
We opt to synthesize each target frame $\TImage$ by predicting the per-pixel GS primitives (denoted as $\SGaussian = \{(\alpha_j, \mu_j, \Sigma_j, c_j)\}_{j=1}^G$) from the corresponding source frame $\SImage$ using a feed-forward reconstruction model $\reconhead(\cdot)$}
, and rendering the primitives at the target pose $\Tcampose$:
\begin{equation}
    \TImage = \renderer(\SGaussian, \Tcampose \otimes \text{inv}(\Scampose)) ~ \text{with} ~ \SGaussian = \reconhead(\SImage)
\end{equation}
This design choice unifies the reconstruction of static and dynamic scene parts, which is particularly challenging for global reconstruction methods. However, the inherent locality of this design can not guarantee the temporal consistency of neighboring reconstructions. Moreover, the pretrained nature of the feed-forward reconstruction model makes it prone to generalization issues at test-time. Therefore, we introduce a test-time optimization scheme, depicted in~\figref{fig:pipeline}, that adapts the pretrained reconstruction model for each unstable video. We take three key measures. (1) To improve each local reconstruction, we propose multi-view photometric supervision from neighboring frames that explicitly accounts for dynamic objects (cf. \secref{sec:method-multiview-loss}). (2) To instill consistency between local reconstructions within a dilated temporal window, we propose regularization on cross-frame GS primitives matched through dense correspondences from optical flow (cf. \secref{sec:method-window-regularization}). (3) To ensure full-frame stabilization, we use video completion to extrapolate the source frames, before passing them to the reconstruction model to get the extrapolated reconstructions rendered as target frames (cf. \secref{sec:method-scene-extrapolation}). 

\begin{figure}[h]
    \includegraphics[width=0.9\linewidth]{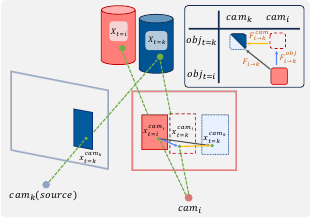}
    \caption{\textbf{Dynamic Object Compensation.} We demonstrate the compensation mechanism that warps dynamic objects at $\text{cam}_i$ to their 'canonical' pose at time $t=k$. Suppose the green point is fixed on a moving cylinder. At time $t=k$, the cylinder is at the blue position and the 3D point $X_{t=k}$ is projected to $x_{t=k}^{\text{cam}_k}$. At time $t=i$, the cylinder has moved to the red position and the 3D point $X_{t=i}$ is projected to $x_{t=i}^{\text{cam}_i}$. To compensate the cylinder's motion that violates the epipolar geometry, we compute conventional optical flow $F_{i\to k}$ and camera motion flow $F_{i\to k}^{\text{cam}}$, and obtain the desired dynamic object flow $F_{i\to k}^{\text{obj}}$ via subtraction. Finally, the solid red rectangle, depicting cylinder at $t=i$ from $\text{cam}_i$, is warped by $F_{i\to k}^{\text{obj}}$ to the dashed red rectangle, depicting cylinder at 'canonical' $t=k$ from $\text{cam}_i$, as if the cylinder did not move at all.}
    \label{fig:compensation}
\end{figure}

\subsection{Scene Extrapolation}
\label{sec:method-scene-extrapolation}

The reconstruction model $\reconhead(\cdot)$ generates per-pixel GS primitives for each source frame. However, depending on the camera motion (in the neighborhood of a source frame) and the target pose, the local reconstruction may lack content required to render a full-frame image at the target pose. Note that, this can be the case despite the aforementioned multi-view supervision and regularization steps. To remedy this situation and ensure full-frame stabilization, we use a video completion method~\cite{zhou2023propainter} to extrapolate the source frames, before passing them to the model to get the extrapolated reconstructions
rendered as target frames (see~\figref{fig:pipeline}). Implementation-wise, a boundary mask is applied on the source frames to indicate the regions to be filled. Then, cross-frame content is propagated by leveraging the optical flow. 

\subsection{Multi-View Dynamics-Aware Photometric Supervision}
\label{sec:method-multiview-loss}

Given a source frame $\SImage$ and its corresponding local reconstruction $\SGaussian$, we use $\SGaussian$ to render images $\SImagepredcus{i} = \renderer(\SGaussian, \Scamposecus{i})$ on temporally adjacent source frames $\SImagecus{i}$ with poses $\Scamposecus{i}$. A photometric loss is consequently used for supervision: 
\begin{equation}
    \cL_{RGB} = \frac{1}{|\neighborimgs|} \sum_{i\in\neighborimgs}
    \left[
    \|\SImagepredcus{i} - \SImagecus{i}\|_1 +
    \lambda_{ssim}\text{SSIM}(\SImagepredcus{i}, \SImagecus{i})
    \right].
\end{equation}
$\neighborimgs$ denotes the set of rendered images consisting of source frame $\SImage$ and $\framesetsize - 1$ randomly sampled neighboring frames within a temporal window ranging $[k - \winsize, k + \winsize]$. To preserve sharpness and details, we employ a combination of L1 and SSIM losses. The set $P_{\neighborimgs}$ contains the corresponding camera poses of rendered images. 

Apparently, any dynamic object violates the epipolar geometry that the photometric loss relies on. To account for this, we propose a compensation mechanism, exemplified in~\figref{fig:compensation}, that warps the dynamic objects in neighboring frames $\SImagecus{i}$ towards their `canonical' pose, i.e. their pose at the \textit{timestamp} of the source frame $\SImagecus{k}$. 

Our compensation is built on decoupling the total 2D pixel motion into camera motion and object motion. 
\pass{We touch the major notations before going through details. $\flow{i}{k}$ stands for the total 2D pixel movement (a.k.a optical flow $\flow{i}{k}=\flowmodel(\SImagecus{i}, \SImagecus{k}$)~\cite{teed2020raft} associated with the time-warping from $t=i$ to $t=k$. $\camflow{i}{k}$ is for the 2D pixel movement only from the ego-camera (a.k.a rigid flow), while $\objectflow{i}{k}$ is for the 2D pixel movement due to object dynamics. 
}

\begin{figure}[h]
    \centering
    \includegraphics[width=0.9\linewidth]{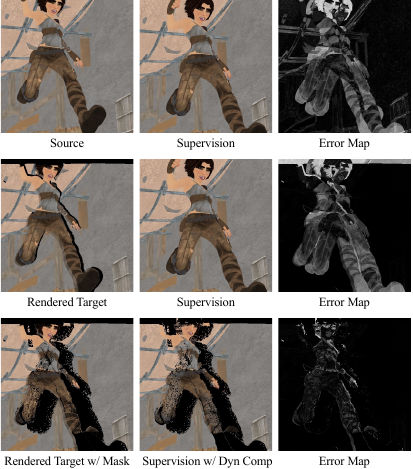}
    \caption{
    \textbf{Demo of Dynamic Compensation in Sintel~\cite{sintel} scene}. 
    \revise{We exemplify our dynamic compensation module with error maps. Ground-truth annotations (e.g. poses, depths and flows) are used for computing the warping. }
    \pass{In the first row we compare source frame $\SImage$ directly with its selected neighboring frame $\SImagecus{i}$, where misalignment comes from both the camera and object motions. In the second row, by rendering the source frame's reconstruction at the neighboring pose, we align static regions but keeps the dynamic misalignment. In the third row, we warp neighboring frames by the proposed flow compensation, effectively reducing errors on the valid region. }
    }
    \label{fig:demo-compensation}
\end{figure}

The dynamic object flow $\objectflow{i}{k}$, can be computed by subtracting from the conventional optical flow $\flow{i}{k}$ by the camera motion flow $\camflow{i}{k}$:
\begin{equation}
\objectflow{i}{k} = \flow{i}{k} - \camflow{i}{k}
\end{equation}
Note that, depth is only known at source frame $\SDepth$, which allows us to compute the `inverse' camera motion flow $\camflow{k}{i}$. To computer the latter, we un-project the 2D pixels of depth map $\SDepth$ into 3D points, apply relative pose change from $\Scampose$ to $\Scamposecus{i}$, then project the 3D points back to 2D pixels, and subtract the original 2D pixel coordinates: 
\begin{equation}
\camflow{k}{i} = \text{Project}_{\camintrinsic, \Scamposecus{i}}(\text{Un-project}_{\camintrinsic, \Scampose}(\SDepth)) - \text{Coordinates}_{k}.   
\end{equation}
In practice, the desired camera motion flow $\camflow{i}{k}$ is generated from the `inverse' camera motion flow $\camflow{k}{i}$ via forward splatting. Further, to reduce erroneous supervision due to occlusion and un-splatted regions, we only supervise and propagate the updates on the valid regions via bi-directional consistency and forward splatting weights. \pass{As indicated in ~\figref{fig:demo-compensation}, our dynamic compensation module effectively reduces the erroneous  supervision from the object motion. Empirically, few warping artifacts exist due to scale change and suboptimal estimation, we can rectify these errors robustly by multi-frame supervision and pretrained priors in neural test-time optimization. }

\subsection{Dilated Reconstruction Regularization}
\label{sec:method-window-regularization}

The multi-view photometric supervision encourages temporal consistency in an implicit way, i.e. by making each local reconstruction as accurate as possible. Admittedly, this is only a half measure. To explicitly impose temporal consistency, we opt to regularize the GS primitives in the temporally adjacent frames to be similar. In particular, given two neighboring reconstructions with per-pixel GS primitives $\SGaussiancus{i}=\reconhead(\SImagecus{i}), \SGaussiancus{j}=\reconhead(\SImagecus{j})$ and their dense correspondences found via bi-directional optical flow $\flow{i}{j}, \flow{j}{i}$, we can define an L2 regularization loss over the matched GS primitives: 
\begin{equation}
    \cL_{\text{pair}}(i,j) = \cM^{i,j} \cdot \parallel \SGaussiancus{i} - \text{warp}(\SGaussiancus{j}, \flow{j}{i}) \parallel_2
\end{equation}
where $\cM^{i,j}$ \pass{indicates the valid regions after bi-directional consistency check. }

\begin{figure}[h]
    \includegraphics[width=0.9\linewidth]{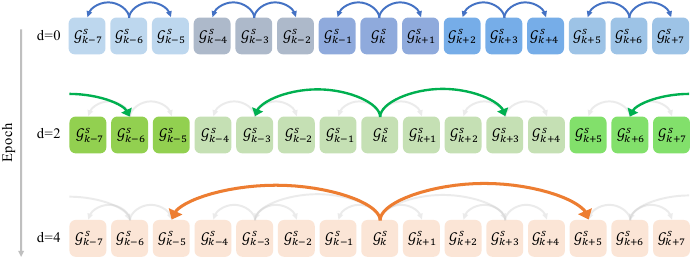}
    \caption{\textbf{Dilated Regularization.} We demonstrate how dilated regularization operates during test time optimization. For simplicity, we set the temporal window $s=3$. Each row shows the GS primitives from consecutive frames. At first epoch, dilation $d=0$, so 3 adjacent reconstructions are regularized. At second epoch, dilation $d=2$, which increases the regularization range up to 9. At third epoch, dilation $d=4$, and regularization range goes up to 27. Note that, these calculations include cumulative effects.}
    \label{fig:dilated}
\end{figure}

As shown in~\figref{fig:dilated}, we bootstrap the regularization within a temporal window of size $s$ with dilation $d$. During test-time optimization, the dilation $d$ is gradually increased to cover a wider temporal range. Mathematically, we compute the regularization as: 
\begin{equation}
    \cL_{\text{consistent}} = \frac{1}{s} \sum_{j=-\left\lfloor s/2 \right\rfloor}^{\left\lfloor s/2 \right\rfloor} \cL_{\text{pair}}(i,~i+j\cdot(d+1)).
\end{equation}

\subsection{Optimization and Motion Smoothing}
\label{sec:optimization}

We optimization the feed forward reconstruction model with total loss: 
\begin{equation}
    \cL = \cL_{\text{RGB}} + \lambda_{\text{consistent}}\cL_{\text{consistent}} + \lambda_{\text{scale}}\cL_{\text{scale}} + \lambda_{\text{offset}}\cL_{\text{offset}}
\end{equation}
where $\cL_{\text{scale}}$ penalizes GS primitives with large scale, and $\cL_{\text{offset}}$ prevents the GS primitives from drifting far away from the re-projected 3D points from depth map \pass{(see supplementary materials for more details).}

To render stabilized frames from the extrapolated reconstructions, following~\cite{lee20213d}, we obtain stabilized camera poses by applying Gaussian filtering to translation and rotation components: 
\begin{equation}
    \Tcampose = \sum_{i=t-w/2}^{t+w/2} g^*(i,k)\Scamposecus{i}
\end{equation}
where $w$ denotes the window size and $g^*(i,k)$ denotes the normalized weights sampled from Gaussian distribution. By adjusting the $\stability$, our system synthesizes videos of different stability levels. 
\begin{equation}
    g(i,k) = e^{-\frac{1}{2}(\frac{|i-k|}{\stability})^2}
\end{equation}

\begin{table*}[h]
    \centering
    \renewcommand{\arraystretch}{1.2}
    \caption{\textbf{Quantitative results on repurposed DeepFused~\cite{shi2022deep} dataset against full-frame baselines.} S: Stability, D: Distortion, GC-S: Sparse Geometry Consistency, GC-D: Dense Geometry Consistency. \textbf{Bold} indicates the best result. We omit the cropping ratio metric, as all methods are full-frame. Overall, our method achieves the best geometry consistency and stability metrics, while maintaining comparable performance on the conventional distortion metric. Interpolation-based DIFRINT~\cite{choi2020deep} fails to generalize and produces fatal artifacts. For scenes with intense camera motions and dynamics, MetaStab~\cite{ali2024harnessing} fails to model complex motions and alleviate instability. Warping-based FuSta~\cite{liu2021hybrid} achieves comparable stability, but at the cost of geometry consistency. }
    \resizebox{\textwidth}{!}
    {%
        \begin{tabular}{lc|ccc|ccc|ccc|cccc}
        &&\multicolumn{3}{c|}{Mild}  & \multicolumn{3}{c|}{Intense}& \multicolumn{3}{c|}{Dynamics} & \multicolumn{4}{c}{Mean} \\
        Method   & Type & S$\uparrow$ & \pass{GC-S}$\downarrow$ & \pass{GC-D}$\uparrow$ & S$\uparrow$ & \pass{GC-S}$\downarrow$ & \pass{GC-D}$\uparrow$ & S$\uparrow$ & \pass{GC-S}$\downarrow$ & \pass{GC-D}$\uparrow$ & D$\uparrow$ & S$\uparrow$ & \pass{GC-S}$\downarrow$ & \pass{GC-D}$\uparrow$ \\
        \hline
        DIFRINT~\cite{choi2020deep}   & 2D & 0.87 & 1.83 & 21.71 & 0.79 & 1.71 & 21.30 & 0.75 & 1.73 & 21.50 & 0.52 & 0.8 & 1.78 & 21.53\\
        FuSta~\cite{liu2021hybrid}    & 2D   &  \firstcolor{0.93} & 0.75 & 29.38 & \secondcolor{0.87} & 1.07 & 26.30 & \secondcolor{0.93} & 0.88 & 26.81 & 0.97 &\secondcolor{0.91} & 0.90 & 27.21 \\
        MetaStab~\cite{ali2024harnessing} & 2D & \secondcolor{0.92} & \secondcolor{0.53} & 30.72 & 0.79 & \secondcolor{0.61} & 29.50 & 0.91 & \secondcolor{0.50} & 29.13 & \firstcolor{0.99} & 0.87 & 0.55 & 29.79  \\
        \hline
        GaVS~($\stability=4$)   & 3D & \firstcolor{0.93} & \firstcolor{0.52} & \firstcolor{32.31}  & 0.85 & \firstcolor{0.60} & \firstcolor{31.22} & \secondcolor{0.93} & \firstcolor{0.49} & \firstcolor{30.28} & \secondcolor{0.98} & 0.90 & \firstcolor{0.53} & \firstcolor{31.27} \\
        GaVS~($\stability=8$)   & 3D & \firstcolor{0.93} & \secondcolor{0.53} & \secondcolor{32.26} & \firstcolor{0.89} & \secondcolor{0.61} & \secondcolor{30.55} & \firstcolor{0.94} & \secondcolor{0.50} & \secondcolor{30.16} & 0.97 & \firstcolor{0.92} & \secondcolor{0.54} & \secondcolor{30.99}
        \end{tabular}
    }    
    \label{tab:full-frame-quantitative}
\end{table*}
\begin{table}[h]
    \centering
    \renewcommand{\arraystretch}{1.2}
    \caption{\textbf{Quantitative results on repurposed DeepFused~\cite{shi2022deep} against cropping-based baselines.} Our methods achieves superior performance on all metrics.}
    \resizebox{1.0\linewidth}{!}
    {%
        \begin{tabular}{lc|ccccc}
            Method       & Type & CR$\uparrow$           & D$\uparrow$             & S$\uparrow$             & \pass{GC-C}$\downarrow$ & \pass{GC-D}$\uparrow$            \\
            \hline
            L1~\cite{grundmann2011auto}           & 2D   & 0.65         & 0.83          & 0.91          & 1.18 & 27.09         \\
            Yu~\cite{yu2020learning} & 2D   & 0.88         & 0.95          & 0.91          & 0.93 & 29.67         \\
            Deep3D~\cite{lee20213d}       & 2.5D & 0.91         & 0.89          & 0.90          & 0.72 & 22.78         \\
            \hline
            GaVS~($\stability=4$)       & 3D   & \textbf{1.0} & \textbf{0.98} & 0.90          & \textbf{0.54} & \textbf{31.27}\\
            GaVS~($\stability=8$)       & 3D   & \textbf{1.0} & 0.96          & \textbf{0.92} & 0.55  & 30.99     
        \end{tabular}
    }
    \label{tab:quantitative_crop}
\end{table}

\section{Experiments}

\subsection{Experimental Setting}
\label{sec:experimental_setting}

\noindent \textbf{Implementation Details.}
Our reconstruction module builds upon Flash3D~\cite{szymanowicz2024flash3d} and UniDepth~\cite{piccinelli2024unidepth}. The decoder is an MLP that predicts per-pixel GS primitives in a two-layered fashion. The use of two-layered predictions is to better model the 3D scene around occlusions, with a different set of GS primitives. \pass{Prior to fine-tuning, we preprocess the input videos using specialized tools: rolling shutter removal \revise{of using shutter and gyro logs to correct frame distortions captured with asynchronous row scanning in capturing}, GLOMAP~\cite{pan2025global} for 3D pose estimation, Grounded SAM~\cite{ren2024grounded} for foreground mask generation, ProPainter~\cite{zhou2023propainter} for video outpainting, and RAFT~\cite{teed2020raft} for optical flow estimation. \revise{Further implementation details can be found in the supplementary materials.} }

During finetuning, the decoder is updated over 3 epochs to balance efficiency and performance. For multi-view supervision, the frame window is configured with $W=10$ views. For the dilated regularization, we use a frame window of $s=5$ and an incremental dilation of $d=[0, 2, 4]$. As our method supports controllable stability, we use $\stability=4$ and $\stability=8$ for evaluation. Experiments are conducted on a single RTX 5000 GPU.


\noindent \textbf{Datasets}
We utilize the gyro-based video stabilization dataset of~\cite{shi2022deep}, which includes diverse camera motions, characterized by varying levels of shakiness, and complex scene dynamics. The videos are categorized into three groups, two for static scenes with different degrees of shakiness, and one for scenes containing dynamic objects: \textit{Mild}, \textit{Intense}, and \textit{Dynamics}. Each group has five samples.

\noindent \textbf{Baselines and Metrics.}
We compare against baselines with publicly available code and/or pretrained models. That is, image-based 2D and 2.5D baselines, including warping-based approaches (L1~\cite{grundmann2011auto}, Yu and Ramamoorthi~\cite{yu2020learning}, FuSta~\cite{liu2021hybrid}, Deep3D~\cite{lee20213d}) and interpolation-based approaches (DIFRINT~\cite{choi2020deep}, MetaStab~\cite{ali2024harnessing}). Additionally in the supplementary file, we compare against a gyro-based 3D baseline (DeepFused~\cite{shi2022deep}).

For quantitative evaluation, we follow prior works and use the metrics: \textbf{\textit{(1) Cropping Ratio}}: measures the valid frame area after cropping off invalid regions. \textbf{\textit{(2) Distortion}}: measures the global anisotropic homography scaling between source and target frame. \textbf{\textit{(3) Stability}}: measures the smoothness of the tracked features. 
\pass{\textbf{\textit{(4) Geometry Consistency}}: Inspired by~\cite{li2024sora}, we introduce Sparse and Dense Geometry Consistency (denoted as GC-S and GC-D) as additional metrics to evaluate local distortions and temporal geometry consistency. }
%
We hypothesize that better geometry consistency results in lower point cloud projection errors during the sparse 3D reconstruction. Furthermore, we verify the per-pixel discrimination on the dense 3D reconstruction to examine visual quality. 
%
To evaluate sparse geometry consistency (GC-C), we utilize GLOMAP~\cite{pan2025global} as a probing tool and report the reprojection error from its statistics. 
\pass{For dense geometry consistency (GC-D), we employ 3DGS~\cite{kerbl20233d} for per-video dense reconstruction. We compute PSNR as GC-D and the test sets are selected with equal intervals of 8 images which is same as the common novel view synthesis benchmarks. We mask out dynamic objects in the reconstruction and evaluation. }

Additionally, we conduct a \textit{user study} to compare our method with other state-of-the-art approaches. Participants were asked to evaluate stabilized video results. For more details, refer to the survey questionnaire provided in the supplementary material.

\subsection{Quantitative Evaluation}
\label{sec:quantitative_evaluation}

As shown in~\tabref{tab:quantitative_crop}, our approach outperforms cropping baselines across all metrics, especially in reducing distortion and preserving geometry consistency. Deep3D~\cite{lee20213d} exploits epipolar regularization for reducing spatial distortions, but it lacks 3D geometry prior and dynamic modeling which prevents its from reaching the optimal. While cropping baselines achieve comparable stability, they suffer from visible distortions and significant content loss due to cropping, hence compromising realism.

We also compare against full-frame baselines in~\tabref{tab:full-frame-quantitative}. \pass{The GC-S and GC-D metric highlights our method's strength on reducing distortions and maintaining consistency, outperforming others across all subsets.} While interpolation-based methods, like MetaStab~\cite{ali2024harnessing}, struggle with complex camera motions and dynamics, and warping-based methods, like Fusta~\cite{liu2021hybrid}, trade-off geometry consistency for stability, our method maintains the best balance between stability and distortion in diverse scenarios.

Having provided these numbers, we would like to stress that the majority of the quantitative metrics only rely on sparse image features, while video stabilization is inherently a dense video synthesis task. Thus, qualitative evaluations, especially video side-by-side comparisons, are better suited to give the full picture.

\begin{figure}[t]
    \centering
    \includegraphics[width=0.95\linewidth]{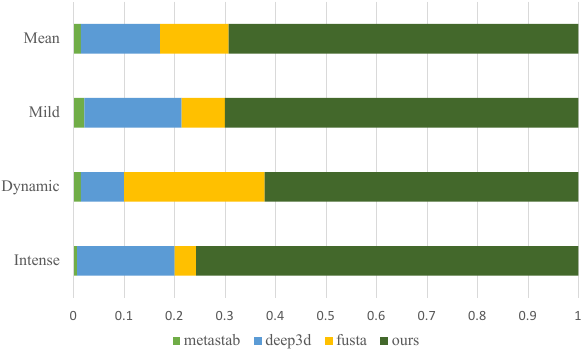}
    \caption{\textbf{User study preference.} We asked 28 participants to evaluate our method, MetaStab~\cite{ali2024harnessing}, Fusta~\cite{liu2021hybrid}, and Deep3D~\cite{lee20213d} across 15 video side-by-side comparisons, and declare their preference with respect to cropping ratio, distortion, stability, and geometric consistency combined. Our method was chosen over 60\% of time, regardless of the camera motion or scene dynamics group the video belongs to.}
    \label{fig:user-study}
\end{figure}

\subsection{Qualitative Evaluation}
\label{sec:qualitative_evaluation}



\begin{figure*}[p]
    \centering
    \includegraphics[width=\textwidth]{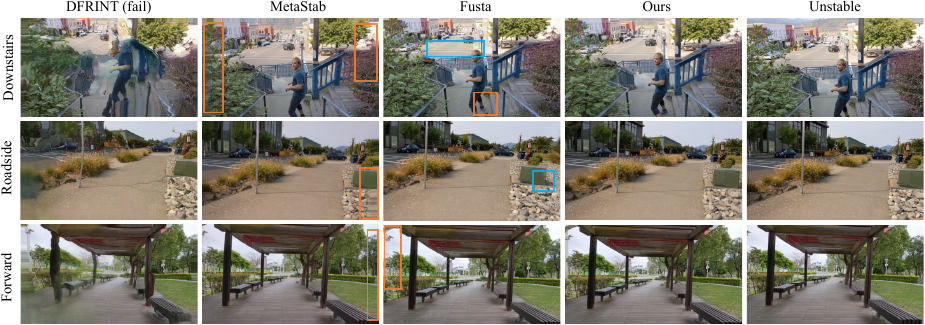}
    \caption{\textbf{Qualitative results against full-frame baselines.} In the first two columns, we use \textcolor{orange}{orange boxes} to indicate blurry outpainting instances and \textcolor{cyan}{cyan boxes} to indicate distortion artifacts. We omit DIFRINT~\cite{choi2020deep} from comparison due to fatal failure. Interpolation-based methods (i.e. MetaStab~\cite{ali2024harnessing}) either fail to generalize to novel scenes or produce blurry outpainting near the boundary. Warping-based methods (i.e. Fusta~\cite{liu2021hybrid}) generate distortions and shearing over time, causing temporal jumping artifacts. Our method achieves high-fidelity rendering with fewer distortions and sharp content on the extrapolated boundary. \\
    \textbf{Important note:} As the aforementioned visual artifacts are hard to be properly depicted in single-frame snapshots, we refer the reader to the supplementary material, where each video can be visually inspected in high resolution, making it easier to spot them.}
    \label{fig:qualitative-full-frame}
\end{figure*}

In~\figref{fig:qualitative-full-frame}, we qualitatively compare against full-frame baselines, MetaStab~\cite{ali2024harnessing}, FuSta~\cite{liu2021hybrid}, and DIFRINT~\cite{choi2020deep}. Our method shows the fewest inconsistency and distortion artifacts with respect to the original content, achieving the most balanced video stabilization performance. 

\begin{figure*}[p]
    \centering
    \includegraphics[width=\textwidth]{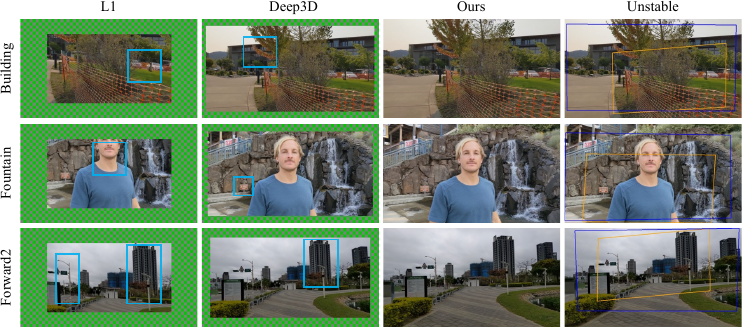}
    \caption{\textbf{Qualitative results against baselines with cropping.} In the unstable frames (last column), we plot bounding boxes to indicate the crop window of the baselines, i.e. \textcolor{orange}{orange box} for L1~\cite{grundmann2011auto} and \textcolor{blue}{blue box} for Deep3D~\cite{lee20213d}. In the first two columns, we use \textcolor{cyan}{cyan boxes} to indicate distortions. All cropping baselines produce visible distortions and excessive cropping in scenes with large camera motions in order to maintain stability. In contrast, our method generates full-frame stabilized videos, while simultaneously producing fewer distortions. \\
    \textbf{Important note:} As geometric distortions are hard to be properly depicted in single-frame snapshots, we refer the reader to the supplementary material, where each video can be visually inspected in high resolution, making it easier to spot such visual artifacts.}
    \label{fig:qualitative-crop}
\end{figure*}

In~\figref{fig:qualitative-crop}, we qualitatively compare against baselines with cropping, L1 Stabilizer~\cite{grundmann2011auto} and Deep3D~\cite{lee20213d}. These works often struggle to maintain geometric consistency, resulting in significant distortions that degrade the user experience. Leveraging the scene extrapolation module, our method achieves full-frame stabilization, without loss of content from the source video, while simultaneously producing fewer distortions. 



\subsection{User Study}
\label{sec:user_study}

We conduct a user study comparing our stabilized videos against state-of-the-art baselines. The study participants were presented with an unstable video and four stabilized results (in randomized order) and asked to select the best overall outcome. The questionnaire is provided in the supplementary material. As shown in~\figref{fig:user-study}, 28 participants evaluated 15 video clips, grouped by the same groups as in the quantitative evaluation. Our method was consistently preferred, especially in challenging scenes with intense camera motion and scene dynamics, where MetaStab~\cite{ali2024harnessing} and Fusta~\cite{liu2021hybrid} produce wobbling artifacts and geometric distortions respectively, while Deep3D~\cite{lee20213d} has excessive cropping. 

\begin{figure}[t]
    \includegraphics[width=0.9\linewidth]{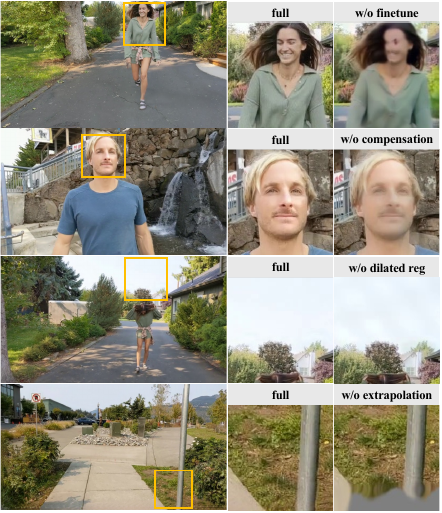}
    \caption{\textbf{Ablation study.} We omit different components of our pipeline (from top to bottom: test-time optimization, dynamics compensation, dilated regularization, scene extrapolation), and study the impact on the results. Clearly, each omitted component has a large impact on visual quality.}
    \label{fig:ablation_study}
\end{figure}

\subsection{Comparison with Global 4D Reconstruction}

\begin{figure}
    \centering
    \includegraphics[width=0.9\linewidth]{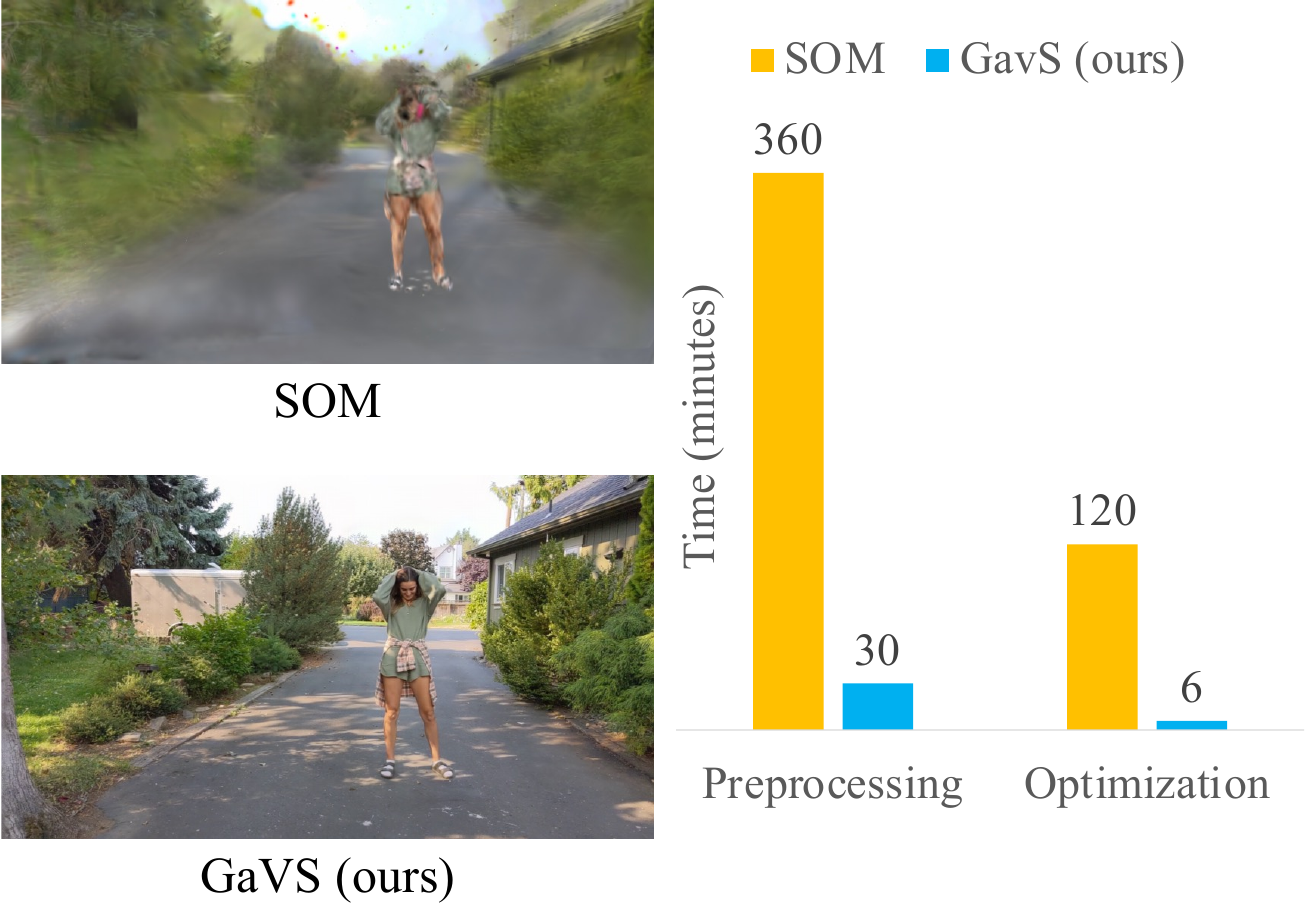}
    \caption{\textbf{Comparison with Global 4D Reconstruction.} 
    \pass{Our method is leading a large magnitude in efficiency on both preprocessing and optimization. Additionally, shape-of-motion~\cite{wang2024shape} fails to reconstruct sharp contents due to poor 2D global tracking and inaccurate 3D poses. }
    }
    \label{fig:global_cmp}
\end{figure}

\pass{We further validate our tailored feed-forward local reconstruction for video stabilization through a preliminary comparison with global 4D reconstruction methods. Specifically, we adopt Shape-of-Motion \cite{wang2024shape} as the baseline, as it employs the same 3DGS representation and offers state-of-the-art efficiency. The comparison is conducted on the \textit{dynamic dance} scene of 455 frames. As shown in \figref{fig:global_cmp}, our method achieves a 12× and 20× speedup in the preprocessing and optimization stages, respectively. Moreover, the global 4D methods are highly dependent on the accuracy of pixel-wise tracking, making them susceptible to failure in the scenes with complex dynamics (e.g., human dancing), where they often produce suboptimal renderings.} 
\revise{Overall, while global reconstruction methods benefit from deformation fields for rendering large baseline changes, the inefficiency and instability of learning these fields under monocular capture make them less practical for video stabilization, where our deformation-free, feed-forward local reconstruction offers a more efficient and robust solution tailored to the small baseline scenarios in this task.}

\subsection{Ablation Study}
\label{sec:abalation_study}

In~\figref{fig:ablation_study}, we visually\footnote{As stabilization metrics are defined on sparse image features, a quantitative ablation fails to capture dense image effects, clearly depicted in the provided visual examples.} analyze the impact of different design choices in our pipeline using the dance (Dynamics), fountain (Dynamics), and roadside (Intense) scenes.

\noindent \textbf{w/o test-time optimization:} Disabling finetuning and relying solely on the pretrained feed-forward Flash3D model~\cite{szymanowicz2024flash3d} results in poor generalization, loss of details, and blurry outputs with unpleasant artifacts.

\noindent \textbf{w/o dynamics compensation:} Test-time optimization without compensation for dynamic objects degrades the reconstruction quality, causing blurriness and loss of textural details, particularly on the man's face, leading to unrealistic rendering.

\noindent \textbf{w/o dilated regularization:} Removing dilated cross-frame regularization naturally weakens the GS primitives consistency, introducing artifacts in textureless regions due to unconstrained scaling and orientation in the GS primitives.

\noindent \textbf{w/o scene extrapolation:} Scene extrapolation from the video completion model significantly improves reconstruction on the boundaries and well preserves the original field of view.

\section{Conclusion and Limitations}

In this paper, we introduced a novel perspective for video stabilization grounded in 3D reconstruction. By operating on a frame-by-frame basis, we feed-forward reconstruct 3D snapshots of the scene, in a unified manner for static and dynamic parts. To handle the inherent inconsistencies of this design choice, we leverage multi-view supervision, robust to dynamics, and regularization across time. The latter happens with an efficient test-time optimization that minimizes domain shift and enhances geometric consistency. Frame outpainting, prior to reconstruction, ensures full-frame stabilization. Numerous evaluations (quantitative, qualitative, user study, ablation) verify that against the state-of-the-art in video stabilization our work achieves the best balance between stability, distortion, crop ratio, and geometry consistency.   

We outline the limitations as follows:

\noindent \textbf{Reliance on visual odometry}: Our 3D-grounded stabilization requires accurate 3D camera poses in local temporal windows. In this work, we use GLOMAP~\cite{pan2025global} for visual odometry, which fails to scale in scenarios such as under-exposed/low-fidelity videos or selfie videos with few static cues. \pass{Additionally, since global SfM methods depend heavily on parallax effects from translational motion, our approach may be more sensitive to errors in the scenarios dominated by rotations. }
A promising direction is robust odometry from multiple sensors, e.g. IMU, depth, which would enable our method to handle such scenarios. 

\noindent \textbf{Trajectory smoothing without constraints}: \pass{Our current stabilization method works in an open-loop way: frames are inpainted at a fixed scale, and camera motion is smoothed separately using a Gaussian filter. However, with very strong motion, the reconstruction might not cover the whole frame. This issue could be improved by optimizing the camera path with field-of-view constraints.}



\noindent \textbf{Failure of reconstruction model}: We finetune Flash3D~\cite{szymanowicz2024flash3d} using UniDepth~\cite{piccinelli2024unidepth} for depth estimation. While test-time optimization corrects minor depth errors, it cannot recover from UniDepth failures. A better depth estimator could address this limitation.

\noindent \textbf{Higher resolutions}: Our per-pixel GS primitives scale poorly with resolution, leading to high GPU demands. More efficient models—such as hierarchical reconstruction or pruning—should be explored.

\newpage

\bibliographystyle{ACM-Reference-Format}

\bibliography{references/references}
\clearpage

\clearpage
\setcounter{page}{1}
\twocolumn[
    \begin{center}
        {\LARGE \textbf{Supplementary Material}}
        \vspace{5mm}
    \end{center}
]
\appendix

In this \textbf{supplementary document}, we compare our method with gyro-based methods, and present more details of regularization, data preprocessing, questionaire design of user study, and per-scene quantitative results. 


\section{Optimization Regularization}

\pass{In the total loss of optimization, we introduce two auxiliary regularization:} $\cL_{\text{scale}}$ and $\cL_{\text{offset}}$. 

$\cL_{\text{scale}}$ penalizes the predicted 3DGS primitives from being too large. Unlike Flash3D~\cite{szymanowicz2024flash3d} with fixed values, we adaptively compute the penalization threshold by incorporating the prior monocular depths and image resolution. Mathematically, we compute $\cL_\text{scale}$ as, 

\begin{equation}
    \cL_{\text{scale}} = \frac{
    \text{Sum}((\Sigma_{\text{gs}} \otimes [\Sigma_{\text{gs}} > \scaleLarge]))}
    {\text{Sum}([\Sigma_{\text{gs}} > \scaleLarge])}
\end{equation}

\begin{equation}
    \scaleLarge = \frac{70 * \text{w}}{\cD}
\end{equation}

where $\Sigma_{\text{gs}}$ refers to the scaling matrix of Gaussian primitives, $\cD$ is the prior monocular depth, $w$ is the finetuning image width and $\otimes$ is the element wise multiplication.

For $\cL_{\text{offset}}$, by regularizing the rendered depth being similar as the prior depth, we prevent the primitives from drifting far away. 

\begin{equation}
    \cL_{\text{offset}} = \frac{\text{Sum}(|\cD_\text{render} - \cD|\otimes [|\cD_\text{render} - \cD| > \scaleDepth])}
    {\text{Sum}(|\cD_\text{render} - \cD| > \scaleDepth)}
\end{equation}

\begin{equation}
    \scaleDepth = 0.2*\cD
\end{equation}

In practice, we found that $\mathcal{L}_{\text{scale}}$ is optional in most cases, while $\mathcal{L}_{\text{offset}}$ performs better when applied exclusively to foreground objects, where epipolar supervision is weaker.

\section{Comparison with Gyro-based Approach}
We compare our method with gyro-based video stabilization method DeepFused~\cite{shi2022deep} where our 3D video stablization datasets are built from. As the method is learning-based and the training data overlaps with some of our test scenes, for fair comparison, we only evaluate on the scenes belonging to the original test scenes. 

As shown in ~\tabref{tab:deep_fused_comparison}, our gryo-free method achieves similar stabilization compared to gyro-based method and introduce fewer distortions. In terms of reprojection errors, both method maintain absolutely low scores indicating good 3D geometry consistency. While sharing similar performance, DeepFused results in noticeable cropping while our method could maintain the original field of view. 



\begin{table}[!htp]
    \centering
    \renewcommand{\arraystretch}{1.2}
    \caption{\textbf{Quantitative Results with DeepFused.} \textbf{Bold} indicates the best. Our gyro-free method shows comparable stability with gyro-based DeepFused method while providing original field of view and introducing fewer distortion. The high GC-S and GC-D scores indicate that both methods well preserve geometry consistency across time.}
    \resizebox{1.0\linewidth}{!}
    {
        \begin{tabular}{l|ccccc}
        Method     & CR $\uparrow$ & D $\uparrow$ & S$\uparrow$ & GC-S$\downarrow$ & GC-D$\uparrow$\\
        \hline
        DeepFused~\cite{shi2022deep}     &  0.83 & 0.94 & 0.92 & \textbf{0.47} & 30.17\\
        GaVS~($\stability=4$)        &  \textbf{1.0} & \textbf{0.97} & 0.92 & 0.48 & \textbf{31.04}    \\
        GaVS~($\stability=8$)   &  \textbf{1.0} & 0.95 & \textbf{0.93} & 0.50 & 30.92
        \end{tabular}
    }
    \label{tab:deep_fused_comparison}
\end{table}

\section{Data Preprocessing}

In this section, we elaborate the implementation of data preprocessing tasks including: \textit{rolling shutter removal}, \textit{camera poses estimation and alignment}, \textit{dynamic object segmentation}, \textit{video completion} and \textit{optical flow}. Additionally, we also report the timing respectively. 

\subsection{Rolling Shutter Removal}

For 3D reconstruction tasks, removing rolling shutter effects is essential to ensure accurate results. RS distortions can disrupt multi-view geometry, leading to errors in camera pose estimatio. Empirically, we found the 3D video stabilization has high requirement for accurate camera estimation. If the rolling shutter effect is preserved, the accumulated error would cause noticeable artifacts across the frames. 

As shown in the Algorithm.~\ref{algo:rs_ois_rm}, we remove the rolling shutter effect from the dataset by utilizing precise gyroscope and OIS data to set the target camera orientation to an average rotation and zero OIS offset. The frame is then divided into multiple row blocks, and a relative 2D homography is calculated between the mean camera position and each block-wise camera. Given the sparse grid with vertices indicating local homography, the final frame is generated by fitting the grid with triangles and computing a dense warping field to sample pixels from the raw image.

\begin{algorithm*}
\caption{Rolling Shutter and OIS Removal}
\label{algo:rs_ois_rm}
\begin{algorithmic}[1]
\Procedure{Remove}{$SrcFrame, GyroHistory, OISHistory, BlockSize$}
    \State $DstExt \gets \texttt{InterpolateMeanRotation}(GyroHistory)$
    \State $DstInt \gets \texttt{ZeroOISCamera}()$
    \State $Step \gets \frac{SrcFrame.Height}{BlockSize}$
    \State $SrcGrid \gets \texttt{MeshGrid}(Step)$
    \State $DstGrid \gets \texttt{ZeroGrid}(Step)$
    
    \State \textbf{For Row = 0 \textbf{to} $Step$}
        \State \hspace{1em} $SrcExt \gets \texttt{InterpolateRotation}(GyroHistory, Row)$
        \State \hspace{1em} $SrcOIS \gets \texttt{InterpolateOISOffset}(OISHistory, Row)$
        \State \hspace{1em} $SrcInt \gets \texttt{OISCamera}(SrcOIS)$
        \State \hspace{1em} $Homography \gets DstInt \otimes DstExt \otimes \texttt{int}(SrcExt) \otimes \texttt{inv}(SrcInt)$
        \State \hspace{1em} $DstGrid[\textsc{Row}, :] \gets Homography \otimes SrcGrid[Row, :]$
    \State \textbf{end for}

    \State $WarpField \gets \texttt{DenseWarping}(DstGrid, SrcGrid)$
    \State $StableFrame \gets \texttt{GridSample}(WarpField, SrcFrame)$
    
    \Return $StableFrame$
\EndProcedure
\end{algorithmic}

\end{algorithm*}

\subsection{Camera Pose Estimation and Alignment}

We run GLOMAP~\cite{pan2025global} to obtain camera poses and sparse point clouds. Specifically, we assume single pinhole camera model in the feature extraction and mask out the regions that contain dynamic objects or belong to rolling shutter masks. Sequential matcher is applied to matching features along consecutive video frames. Finally, we run GLOMAP with default configuration. 

Next, we estimate a global scale to align the estimated camera poses and predicted dense depth maps from UniDepth~\cite{piccinelli2024unidepth}. For robustness, as shown in the Algorithm.~\ref{algo:depth_alignment}, we use RANSAC for the scale estimation. For each video frame's depth $\SDepth$, we extract the visible point clouds from GLOMAP and generate sparse depth maps. We then use RANSAC to estimate the scale factor $\framescale_k$, aligning the poses towards the depth scale. Then, we robustly determine a global scale by taking the median of the scale sequence, $\framescale^* = \text{median}(\{\framescale\})_{k=1}^T$. This alignment provides a reliable initialization, which improves the stability and efficiency of the subsequent reconstruction fine-tuning process. 

\begin{algorithm*}
\begin{algorithmic}[1]
\Procedure{ScaleAlignment}{$MetricDepth, CamPose, SparsePts$}

\State $SparseDepth \gets \texttt{project}(CamPose, SparsePts)$
\State $Mask \gets \texttt{where}(SparseDepth)$ \Comment{image-shape mask, true if depth exists}

\State $SparseMetricDepth \gets MetricDepth[Mask]$

\State $LogSparseDepth \gets \texttt{log}(SparseDepth)$
\State $LogSparseMetricDepth \gets \texttt{log}(SparseMetricDepth)$ \\ \Comment{in log scale, near and far errors have different weights}

\State \textbf{RANSAC}
    \State \hspace{1em} $Sample \gets \texttt{RandomSample}(Mask) $
    \State \hspace{1em} $LogScale \gets (LogSparseDepth[Sample] - LogSparseMetricDepth[Sample])$
    \State \hspace{1em} $Scale \gets LogScale.mean().exp()$  \Comment{compute scale from a subset}
    \State \hspace{1em} $Inliers \gets \texttt{abs}(LogSparseDepth - LogSparseMetricDepth\times Scale) < \tau$  \Comment{inlier detection}
    \State \hspace{1em} \textbf{if} {$Inliers > BestInliers$}    \Comment{update the best scale}
        \State \hspace{2em} $BestScale \gets Scale$
        \State \hspace{2em} $BestInliers \gets Inliers$
    \State \hspace{1em} \textbf{endif}
\State \textbf{endfor}

\Return BestScale

\EndProcedure
\end{algorithmic}
\caption{Robust Depth Scale Between Dense Depth and Sparse Reconstruction}
\label{algo:depth_alignment}
\end{algorithm*}

\subsection{Dynamic Object Segmentation}

We extract dynamic masks across video frames using Grounded SAM~\cite{ren2024grounded} which combines Grounding DINO~\cite{ravi2024sam2segmentimages} and SAM2~\cite{ravi2024sam2segmentimages}. It processes a video scene by extracting frames and applying object detection via Grounding DINO with a given text prompt. The detected object bounding boxes are then used to generate segmentation masks with SAM 2. In our implementation, we provide the common dynamic classes such as \textit{vehicle} and \textit{human} as prompts for segmentation. 

\subsection{Video Completion}

We utilize out-painted video frames to perform reconstruction extrapolation. The out-painting process is carried out using the video completion module ProPainter~\cite{zhou2023propainter}, which first estimates optical flow within the target region and then propagates information from neighboring frames. This propagation is conducted across both the hybrid RGB and feature domains to effectively fill in the missing parts. 
Technically, we add a padding of 96 pixels horizontally and vertically on each frame and use masks to explicitly indicate the regions to be filled. Then, the input video is divided into sub-videos of length 100 frames and an overlapping of 20 frames for optical estimation and image-domain propagation. Next, the frames are propoagated in the feature-level domain with neighboring stride of 15 and reference frames of 10. Finally, the propagated frames are decoded and fused with the original content by padding masks. 

\subsection{Optical Flow}

We use RAFT~\cite{teed2020raft} for computing 2D optical flows. For each frame, we select 20 temporally neighboring frames to compute bidirectional flows as well as occupancy masks. We set the estimation iteration of 20. 

\subsection{Timing}

In ~\tabref{tab:preprocess_time}, we report the timing of our preprocessing tasks in the \textit{dance} scene with 455 frames. We preprocess the data on the server with 4 CPU and 1 RTX 5000 Ada GPU. Assuming enough computing resources, majority of these preprocessing tasks could be executed in parallel, resulting the critical task path of \textit{rolling shutter removal} -> \textit{dynamic mask} -> \textit{Pose Estimation and Sparse Reconstruction}, which takes 13 minutes. Compared to the 4D dynamic reconstruction method~\cite{wang2024shape} which takes >10 hours for preprocessing and 1 hour for optimization, our method achieves 46x and 10x speedup respectively. 

\begin{table}
    \centering
    \caption{\textbf{Timing of Preprocess Tasks.} Test Scene: \textit{dance}, Frame Number: 455}
    \renewcommand{\arraystretch}{1.2}
    \begin{tabular}{l|c}
    \textbf{Preprocessing Task}     & \textbf{Time} \\
    \hline
    Rolling Shutter and OIS Removal    &   1 min \\
    Video Completion & 8 min \\
    Dynamic Mask & < 1min \\
    Pose Estimation and Sparse  & 11 min \\
    Dense Depth & 8 min \\
    Optical Flow & 10 min \\
    \end{tabular}
    \label{tab:preprocess_time}
\end{table}

\section{User Study}

We conduct our user study via online Google form. We list all the fifteen video clips from the repurposed DeepFused~\cite{shi2022deep} dataset in a random order. The original unstable is presented to the participants and the stabilized videos from the four methods (ours, Deep3D, MetaStab and Fusta) are concatenated in a random order and played in synchronization. For the videos with cropping, we resize them to the full-frame resolution with anti-aliasing interpolation. Participants are allowed to pause and replay the videos for thorough evaluation. Finally, they are asked to finish the single-choice question to decide the best overall result. For reference, we provide their evaluation suggestions of visual quality, stability, geometry consistency and content preservation, but ask the participants to make the final trade-off and overall decision. For video quality, participants are also asked to report the device on which they finish the questionnaires and only the results from the computers are taken into consideration. We show one demo question from the \href{https://docs.google.com/forms/d/e/1FAIpQLSeIRDDN8t0SXJVitRDgM2bbGtfCDskURiwo9PgSz-zYTM6Ozg/viewform?usp=header}{\textcolor{blue}{online questionnaire}} as in \figref{fig:user-study}. 

\begin{figure}
    \centering
    \includegraphics[width=0.8\linewidth]{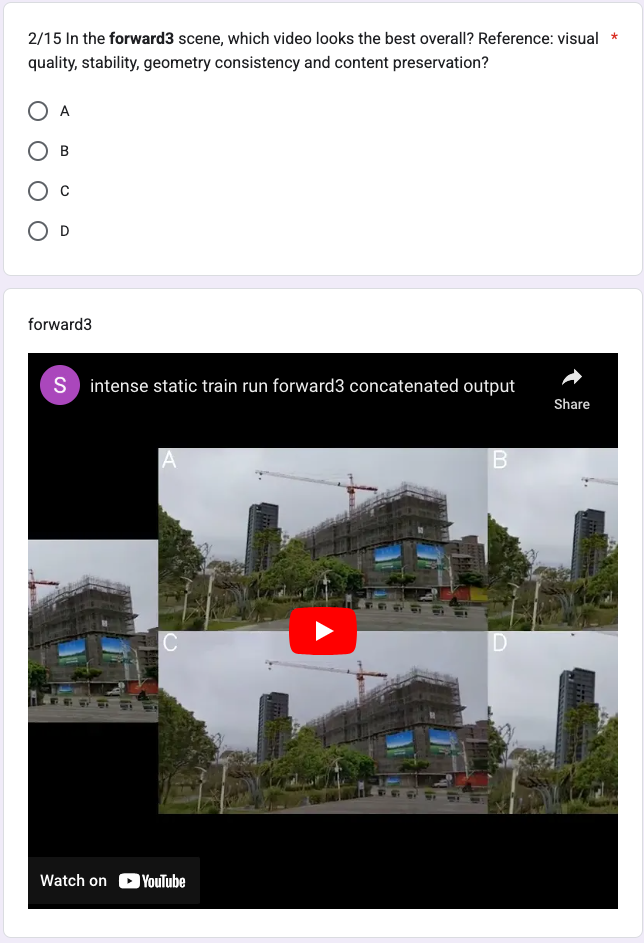}
    \caption{\textbf{Example Question of User Study Survey}}
    \label{fig:user-study}
\end{figure}

\begin{table*}[!]
\centering
\caption{\textbf{Per-scene Distortion on Repurposed DeepFused~\cite{shi2022deep}} dataset. }
\vspace{-4mm}
\resizebox{\textwidth}{!}{
\begin{tblr}{
  column{odd} = {c},
  column{4} = {c},
  column{6} = {c},
  column{8} = {c},
  column{10} = {c},
  cell{2}{1} = {r=5}{},
  cell{7}{1} = {r=5}{},
  cell{12}{1} = {r=5}{},
  vline{3} = {-}{},
  hline{2,7,12,17} = {-}{},
}
                  &            & \textbf{L1}~ & \textbf{FuSta}~ & \textbf{Yu}~ & \textbf{DIFRINT}~ & \textbf{MetaStab}~ & \textbf{Deep3D}~ & \textbf{GaVS} ($\stability$=4) & \textbf{GaVS} ($\stability$=8) \\
\textbf{Mild}     & bush       & 0.89         & 0.96            & 0.93         & 0.48             & 0.99               & 0.94             & 0.93                           & 0.92                           \\
                  & rotation   & 0.90         & 0.99            & 0.98         & 0.54             & 1.00               & 0.97             & 0.99                           & 0.99                           \\
                  & rotation2  & 0.01         & 0.99            & 0.95         & 1.00             & 0.99               & 0.97             & 0.99                           & 0.99                           \\
                  & walk       & 0.89         & 0.99            & 0.99         & 0.47             & 1.00               & 0.97             & 0.99                           & 0.98                           \\
                  & building   & 0.89         & 0.98            & 0.95         & 0.20             & 0.99               & 0.90             & 0.98                           & 0.98                           \\
\textbf{Intense}  & forward    & 0.89         & 0.98            & 0.97         & 0.49             & 0.99               & 0.89             & 0.98                           & 0.96                           \\
                  & forward2   & 0.89         & 0.98            & 0.98         & 0.69             & 0.99               & 0.86             & 0.96                           & 0.95                           \\
                  & forward3   & 0.89         & 0.98            & 0.98         & 0.64             & 0.99               & 0.90             & 0.97                           & 0.96                           \\
                  & slerp      & 0.89         & 0.98            & 0.97         & 0.70             & 0.99               & 0.93             & 0.98                           & 0.97                           \\
                  & roadside   & 0.90         & 0.95            & 0.94         & 0.49             & 0.97               & 0.82             & 0.95                           & 0.92                           \\
\textbf{Dynamics} & dance      & 0.90         & 0.97            & 0.96         & 0.70             & 0.99               & 0.90             & 0.98                           & 0.97                           \\
                  & downstairs & 0.90         & 0.96            & 0.89         & 0.31             & 0.99               & 0.85             & 0.96                           & 0.94                           \\
                  & fountain   & 0.91         & 0.96            & 0.89         & 0.31             & 0.99               & 0.86             & 0.98                           & 0.97                           \\
                  & vehicle    & 0.89         & 0.97            & 0.97         & 0.08             & 0.99               & 0.91             & 0.99                           & 0.97                           \\
                  & walk       & 0.90         & 0.98            & 0.96         & 0.64             & 0.99               & 0.85             & 0.98                           & 0.97                           
\end{tblr}
}
\end{table*}
\begin{table*}[!]
\centering
\vspace{1cm}
\caption{\textbf{Per-scene Stability on Repurposed DeepFused~\cite{shi2022deep}} dataset. }
\vspace{-4mm}
\resizebox{\textwidth}{!}{
\begin{tblr}{
  column{odd} = {c},
  column{4} = {c},
  column{6} = {c},
  column{8} = {c},
  column{10} = {c},
  cell{2}{1} = {r=5}{},
  cell{7}{1} = {r=5}{},
  cell{12}{1} = {r=5}{},
  vline{3} = {-}{},
  hline{2,7,12,17} = {-}{},
}
                  &            & \textbf{L1}~ & \textbf{FuSta}~ & \textbf{Yu}~ & \textbf{DIFRINT}~ & \textbf{MetaStab}~ & \textbf{Deep3D}~ & \textbf{GaVS} ($\stability$=4) & \textbf{GaVS} ($\stability$=8) \\
\textbf{Mild}     & bush       & 0.96         & 0.96            & 0.96         & 0.95             & 0.95               & 0.96             & 0.96                           & 0.96                           \\
                  & rotation   & 0.93         & 0.92            & 0.92         & 0.86             & 0.92               & 0.92             & 0.92                           & 0.92                           \\
                  & rotation2  & 0.95         & 0.95            & 0.95         & 0.83             & 0.95               & 0.95             & 0.95                           & 0.95                           \\
                  & walk       & 0.92         & 0.91            & 0.92         & 0.89             & 0.90               & 0.92             & 0.92                           & 0.93                           \\
                  & building   & 0.92         & 0.91            & 0.89         & 0.82             & 0.88               & 0.89             & 0.90                           & 0.90                           \\
\textbf{Intense}  & forward    & 0.92         & 0.90            & 0.90         & 0.94             & 0.89               & 0.90             & 0.89                           & 0.90                           \\
                  & forward2   & 0.95         & 0.89            & 0.93         & 0.93             & 0.88               & 0.92             & 0.91                           & 0.92                           \\
                  & forward3   & 0.77         & 0.81            & 0.79         & 0.70             & 0.66               & 0.80             & 0.73                           & 0.82                           \\
                  & slerp      & 0.87         & 0.81            & 0.83         & 0.80             & 0.64               & 0.82             & 0.77                           & 0.86                           \\
                  & roadside   & 0.90         & 0.96            & 0.89         & 0.57             & 0.90               & 0.90             & 0.95                           & 0.97                           \\
\textbf{Dynamics} & dance      & 0.94        & 0.94           & 0.90                 & 0.49            & 0.91              & 0.94            & 0.93                           & 0.94                            \\
                  & downstairs & 0.88         & 0.89            & 0.88         & 0.78             & 0.87               & 0.90             & 0.89                           & 0.91                           \\
                  & fountain   & 0.96         & 0.95            & 0.95         & 0.65             & 0.94               & 0.95             & 0.95                           & 0.96                           \\
                  & vehicle    & 0.94         & 0.96            & 0.96         & 0.90             & 0.94               & 0.94             & 0.95                           & 0.95                           \\
                  & walk       & 0.82         & 0.92            & 0.92         & 0.92             & 0.91               & 0.92             & 0.91                           & 0.92                           
\end{tblr}
}
\end{table*}

\begin{table*}[!]
\centering
\caption{\textbf{Per-scene Cropping Ratio on Repurposed DeepFused~\cite{shi2022deep}} dataset. }
\vspace{-4mm}
\resizebox{1.0\textwidth}{!}{
\begin{tblr}{
  column{odd} = {c},
  column{4} = {c},
  column{6} = {c},
  column{8} = {c},
  column{10} = {c},
  cell{2}{1} = {r=5}{},
  cell{7}{1} = {r=5}{},
  cell{12}{1} = {r=5}{},
  vline{3} = {-}{},
  hline{2,7,12,17} = {-}{},
}
                  &            & \textbf{L1} & \textbf{Fusta} & \textbf{Yu} & \textbf{DIFRINT} & \textbf{MetaStab} & \textbf{Deep3D} & \textbf{GaVS} ($\stability$=4) & \textbf{GaVS} ($\stability$=8) \\
\textbf{Mild}     & bush       & 0.64        & 1.00           & 0.84                 & 1.00            & 1.00              & 0.91            & 1.00                           & 1.00                           \\
                  & rotation   & 0.64        & 1.00           & 0.95                 & 1.00            & 1.00              & 0.97            & 1.00                           & 1.00                           \\
                  & rotation2  & 0.63        & 1.00           & 0.93                 & 1.00            & 1.00              & 0.96            & 1.00                           & 1.00                           \\
                  & walk       & 0.64        & 1.00           & 0.94                 & 1.00            & 1.00              & 0.96            & 1.00                           & 1.00                           \\
                  & building   & 0.65        & 1.00           & 0.93                 & 1.00            & 1.00              & 0.95            & 1.00                           & 1.00                           \\
\textbf{Intense}  & forward    & 0.64        & 1.00           & 0.90                 & 1.00            & 1.00              & 0.93            & 1.00                           & 1.00                           \\
                  & forward2   & 0.64        & 1.00           & 0.87                 & 1.00            & 1.00              & 0.91            & 1.00                           & 1.00                           \\
                  & forward3   & 0.66        & 1.00           & 0.83                 & 1.00            & 1.00              & 0.90            & 1.00                           & 1.00                           \\
                  & slerp      & 0.66        & 1.00           & 0.91                 & 1.00            & 1.00              & 0.92            & 1.00                           & 1.00                           \\
                  & roadside   & 0.65        & 1.00           & 0.69                 & 1.00            & 1.00              & 0.76            & 1.00                           & 1.00                           \\
\textbf{Dynamics} & dance      & 0.66        & 1.00           & 0.85                 & 1.00            & 1.00              & 0.93            & 1.00                           & 1.00                           \\
                  & downstairs & 0.65        & 1.00           & 0.87                 & 1.00            & 1.00              & 0.84            & 1.00                           & 1.00                           \\
                  & fountain   & 0.66        & 1.00           & 0.83                 & 1.00            & 1.00              & 0.94            & 1.00                           & 1.00                           \\
                  & vehicle    & 0.64        & 1.00           & 0.92                 & 1.00            & 1.00              & 0.94            & 1.00                           & 1.00                           \\
                  & walk       & 0.65        & 1.00           & 0.92                 & 1.00            & 1.00              & 0.91            & 1.00                           & 1.00                           
\end{tblr}

}
\end{table*}

\begin{table*}[!]
\centering
\vspace{1cm}
\caption{\textbf{Per-scene Sparse Geometry Consistency on Repurposed DeepFused~\cite{shi2022deep}} dataset. }
\vspace{-4mm}
\resizebox{\textwidth}{!}{
\begin{tblr}{
  column{odd} = {c},
  column{4} = {c},
  column{6} = {c},
  column{8} = {c},
  column{10} = {c},
  cell{2}{1} = {r=5}{},
  cell{7}{1} = {r=5}{},
  cell{12}{1} = {r=5}{},
  vline{3} = {-}{},
  hline{2,7,12,17} = {-}{},
}
                  &            & \textbf{L1}~ & \textbf{FuSta}~ & \textbf{Yu}~ & \textbf{DIFRINT}~ & \textbf{MetaStab}~ & \textbf{Deep3D}~ & \textbf{GaVS} ($\stability$=4) & \textbf{GaVS} ($\stability$=8) \\
\textbf{Mild}     & bush       & 0.87         & 0.68            & 0.70         & 1.77             & 0.42               & 0.57             & 0.40                           & 0.41                           \\
                  & rotation   & 0.96         & 0.85            & 0.86         & 1.85             & 0.63               & 0.70             & 0.63                           & 0.63                           \\
                  & rotation2  & 0.99         & 0.72            & 0.75         & 1.89             & 0.48               & 0.61             & 0.50                           & 0.50                           \\
                  & walk       & 1.17         & 0.72            & 0.75         & 1.79             & 0.52               & 0.62             & 0.52                           & 0.53                           \\
                  & building   & 0.96         & 0.76            & 0.74         & 1.84             & 0.61               & 0.69             & 0.56                           & 0.57                           \\
\textbf{Intense}  & forward    & 1.37         & 1.09            & 1.16         & 1.77             & 0.65               & 0.65             & 0.65                           & 0.66                           \\
                  & forward2   & 2.15         & 1.04            & 1.17         & 1.80             & 0.58               & 1.01             & 0.61                           & 0.62                           \\
                  & forward3   & 1.30         & 1.19            & 1.28         & 1.65             & 0.58               & 0.63             & 0.58                           & 0.62                           \\
                  & slerp      & 1.30         & 1.05            & 1.10         & 1.56             & 0.69               & 0.67             & 0.69                           & 0.70                           \\
                  & roadside   & 1.05         & 0.99            & 1.02         & 1.76             & 0.53               & 0.81             & 0.47                           & 0.50                           \\
\textbf{Dynamics} & dance      & 1.23         & 0.86            & 0.83         & 1.85             & 0.40               & 0.67             & 0.41                           & 0.41                           \\
                  & downstairs & 1.17         & 0.89            & 0.90         & 1.74             & 0.55               & 0.98             & 0.54                           & 0.57                           \\
                  & fountain   & 1.13         & 1.12            & 1.14         & 1.60             & 0.55               & 0.92             & 0.53                           & 0.54                           \\
                  & vehicle    & 1.09         & 0.79            & 0.78         & 1.77             & 0.57               & 0.70             & 0.55                           & 0.56                           \\
                  & walk       & 1.03         & 0.76            & 0.80         & 1.71             & 0.40               & 0.48             & 0.42                           & 0.43                           
\end{tblr}

}
\end{table*}

\begin{table*}[!]
\centering
\caption{\textbf{Per-scene Dense Geometry Consistency on Repurposed DeepFused~\cite{shi2022deep}} dataset. }
\vspace{-4mm}
\resizebox{\textwidth}{!}{
\begin{tblr}{
  column{odd} = {c},
  column{4} = {c},
  column{6} = {c},
  column{8} = {c},
  column{10} = {c},
  cell{2}{1} = {r=5}{},
  cell{7}{1} = {r=5}{},
  cell{12}{1} = {r=5}{},
  vline{3} = {-}{},
  hline{2,7,12,17} = {-}{},
}
                  &            & \textbf{L1}~ & \textbf{Fusta}~ & \textbf{Yu}~ & \textbf{DIFRINT}~ & \textbf{MetaStab}~ & \textbf{Deep3D}~ & \textbf{GaVS} ($\stability$=4) & \textbf{GaVS} ($\stability$=8) \\
\textbf{Mild}     & bush       & 27.70        & 27.54           & 29.63        & 22.84            & 29.99              & 24.39            & 31.80                          & 31.95                          \\
                  & rotation   & 27.79        & 26.46           & 29.09        & 19.02            & 27.51              & 21.43            & 27.97                          & 27.95                          \\
                  & rotation2  & 26.94        & 34.12           & 36.04        & 23.27            & 35.00              & 26.51            & 36.64                          & 36.53                          \\
                  & walk       & 28.46        & 30.11           & 32.81        & 22.86            & 30.89              & 23.66            & 32.21                          & 32.03                          \\
                  & building   & 28.37        & 28.70           & 31.36        & 20.54            & 30.23              & 23.18            & 32.95                          & 32.82                          \\
\textbf{Intense}  & forward    & 26.75        & 26.37           & 28.63        & 19.45            & 28.99              & 20.40            & 31.11                          & 31.02                          \\
                  & forward2   & 22.51        & 25.24           & 27.63        & 22.57            & 30.79              & 20.33            & 31.63                          & 31.70                          \\
                  & forward3   & 26.76        & 24.71           & 27.31        & 23.78            & 31.13              & 19.87            & 30.44                          & 30.47                          \\
                  & slerp      & 28.71        & 25.88           & 27.51        & 23.02            & 29.27              & 22.18            & 31.98                          & 29.00                          \\
                  & roadside   & 26.77        & 25.07           & 28.63        & 17.68            & 27.33              & 21.17            & 30.95                          & 30.54                          \\
\textbf{Dynamics} & dance      & 28.80        & 28.53           & 31.57        & 23.71            & 31.45              & 24.18            & 32.55                          & 32.39                          \\
                  & downstairs & 24.14        & 23.85           & 27.50        & 17.24            & 26.44              & 20.18            & 28.65                          & 28.32                          \\
                  & fountain   & 28.80        & 26.54           & 28.60        & 20.52            & 29.03              & 23.09            & 29.91                          & 29.78                          \\
                  & vehicle    & 30.87        & 30.22           & 32.51        & 24.69            & 31.92              & 24.08            & 33.74                          & 33.70                          \\
                  & walk       & 22.96        & 24.91           & 26.28        & 21.79            & 26.84              & 26.92            & 26.57                          & 26.62                          
\end{tblr}

}
\end{table*}

\end{document}